\documentstyle[aps,prl,twocolumn,psfig,floats]{revtex}
\begin{document}
\draft
\wideabs{
\title{Collective Pinning of a Frozen Vortex Liquid in Ultrathin
Superconducting YBa$_2$Cu$_3$O$_7$ Films}
\author{M. Calame$^{1}$, S.E. Korshunov$^{2}$, Ch. Leemann$^{1}$, and P.
Martinoli$^{1}$}
\address{$^{1}$Institut de Physique, Universit\'{e} de Neuch\^{a}tel,
CH-2000 Neuch\^{a}tel, Switzerland}
\address{$^{2}$L.D. Landau Institute for Theoretical Physics,
Kosygina 2, 117940 Moscow, Russia}
\maketitle
\begin{abstract}
The linear dynamic response of the two-dimensional (2D) vortex medium in 
ultrathin YBa$_2$Cu$_3$O$_7$ films was studied by measuring their ac sheet 
impedance $Z$ over a broad range of frequencies $\omega$. With decreasing 
temperature the dissipative component of $Z$ exhibits, at a temperature 
$T^{*}(\omega)$ well above the melting temperature of a 2D 
vortex crystal, a crossover from a thermally activated regime involving 
single vortices to a regime where the response has features consistent 
with a description in terms of a collectively pinned vortex manifold. This 
suggests the idea of a vortex liquid which, below $T^{*}(\omega)$, appears 
to be frozen at the time scales $1/\omega$ of the experiments.  
\end{abstract}
}

It is generally accepted that the effect of weak disorder on flux lattices
in superconductors results in some sort of glassy state \cite{F,FGLV}.
In a rigorous sense, however, the existence of a truly superconducting
vortex glass phase, such that the {\em linear} resistance
vanishes, is impossible in two dimensions \cite{FGL,VKK,Rev}. Although
the random potential associated with disorder can quench the motion of
the vortex medium as a whole, in two dimensions (in contrast to three)
the flow of magnetic flux can be mediated by the motion of thermally created
point-like defects.
To the best of our knowledge,
only limited experimental evidence for the absence of the vortex-glass
transition in two dimensions has been found so far \cite{D,SA}. 

Relying on sheet impedance measurements, in this
Letter we demonstrate that, at typical laboratory time scales, the
linear response of the two-dimensional (2D) vortex medium in ultrathin
YBa$_2$Cu$_3$O$_7$ (YBCO) films to a small driving ac field
drastically changes with decreasing temperature. At high
temperatures the dissipative component of the
impedance is frequency independent and exhibits a strong
(exponential) temperature dependence which can be attributed to
the thermally activated motion of single vortices. However, at
lower temperatures,
this contribution becomes too
small to be observed. In this regime the dynamic response is governed
by a different process whose frequency ($\omega$) and temperature ($T$)
dependences are found to be consistent with the idea of a collectively pinned
vortex medium, in which large portions of the vortex
manifold are fluctuating between pairs of metastable states
in the random potential landscape provided
by the film microstructure \cite {KV,FFH,K}. 

The crossover to the collective pinning regime has been observed
over the whole frequency range explored in our experiments. Quite
remarkably, it is located at a temperature
$T^{*}(\omega)$ which turns out to be much higher than the
estimated melting temperature of a 2D vortex crystal in the
absence of disorder. Thus, the response we observe well below
$T^{*}(\omega)$ does not originate from a collectively pinned
vortex crystal, but from a collectively pinned (dynamically)
frozen vortex liquid (or, in other terms, from a strongly
disordered vortex solid). On the other hand, a study of the
liquid-like response well above $T^{*}(\omega)$ as a function of
the applied magnetic field $B$ reveals that in this regime the
vortex activation process is controlled by a surface barrier
mechanism \cite{B}. 

It is worth mentioning that the classical theory of flux creep in
the collective pinning regime \cite{FGLV,VKK,Rev} and its ac
extensions \cite{KV,FFH,K} ignore the periodic nature of the
vortex lattice, the vortex medium being treated as an elastic
manifold interacting with a random potential. Therefore, this
approach should provide a reliable description of a disordered
vortex solid (or frozen vortex liquid) as long as the frequency of
observation is high enough for the motion of defects with respect
to the driven vortex medium to appear to be frozen. 

A possible criterion to distinguish between a genuine continuous
phase transition to a vortex glass and a simple dynamic crossover
to a collective pinning regime could rely on studies of
$T^{*}(\omega)$ extending at ultralow frequencies. If a true phase
transition occurs, $T^{*}(\omega)$ should saturate with decreasing
$\omega$ at the (nonvanishing) vortex-glass transition
temperature, whereas in the opposite case the decrease of
$T^{*}(\omega)$ should be unlimited. In our experiments
$T^{*}(\omega)$ decreases by only $5\%$ over three
decades in frequency, thereby preventing us from drawing any
conclusion with regard to the existence of a genuine phase
transition. Qualitatively, in the frequency range explored in our
experiments the ac response of ultrathin YBCO layers is identical
to that of thick films, for which signatures of a true 3D vortex-glass 
phase were reported by several authors. Thus, from a practical point 
of view the glass-like features we
observe below $T^{*}(\omega)$ 
turn out to be undistinguishable from those of a genuine
vortex glass. 

The experiments were performed on two films \mbox{YBCO-2} and
YBCO-4, respectively 2 and 4 unit cells thick (\textit{i.e}, with
thicknesses $d=2.4 nm$ and $d=4.8 nm$), grown epitaxially 
($c$-axis oriented) by laser ablation onto ($100$) SrTiO$_{3}$
substrates. Both films were sandwiched between nonsuperconducting buffer 
and cover 
PrBa$_2$Cu$_3$O$_7$ layers. For comparison, a thick ($d=110 nm$) YBCO film 
was also studied. Their complex sheet impedance $Z$ was extracted from 
the mutual-inductance change of a drive-receive coil system \cite{J} operated 
with a conventional lock-in detector allowing inductances to be measured 
with a sensitivity of $\sim10 pH$ between $30 Hz$ and $100 kHz$.
The bulk in-plane magnetic penetration depth $\lambda_{ab}(T)$
was inferred from measurements of the inverse sheet kinetic inductance
$1/L_{0}\equiv\omega\mbox{Im}[1/Z\mbox{$(B=0)$}]=d/\mu_{0}\lambda_{ab}^{2}$
in zero magnetic field and found to fit well,
over a wide temperature range, a parabolic dependence \cite{T}
$\lambda_{ab}^{-2}(T)=\lambda_{ab}^{-2}(0)\mbox{$[1-(T/T_{c})^{2}]$}$ with
$\lambda_{ab}(0)=550 nm$ and $T_{c}=48.3 K$ for YBCO-2 and
$\lambda_{ab}(0)=340 nm$ and $T_{c}=73.7 K$ for YBCO-4.

\begin{figure}
\centerline{\psfig{file=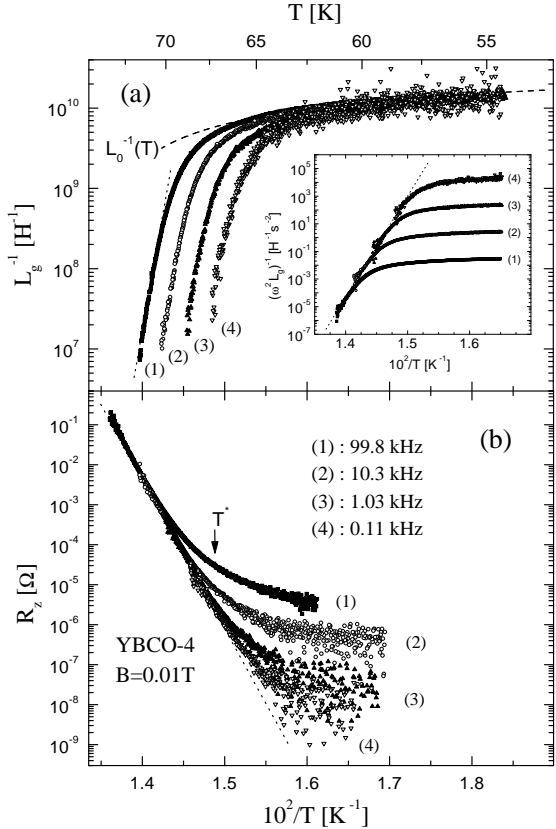,width=0.95\columnwidth,clip=}}
\vspace{1mm}
\caption{Arrhenius plots of ($a$) $1/L_{g}$  
and ($b$) $R_{z}$ of YBCO-4 at $B=0.01T$ for 4 
representative frequencies. The dashed curve in 
($a$) is the inverse kinetic inductance in zero field. $T^{*}$ 
indicates the crossover temperature of $R_{z}$ at $99.8kHz$.}
\label{fig1}
\end{figure}

The inverse sheet inductance $1/L_{g}=\omega\mbox{Im}(1/Z)$
(which measures the degree of superconducting phase coherence in the
system) and the resistive component of the sheet impedance
$R_{z}=\mbox{Re}(Z)$ (which measures dissipation) of YBCO-4
in a perpendicular field of $0.01T$ are plotted logarithmically in Fig.
1 as a function of $1/T$ for a set of 4 representative frequencies. As
highlighted by the dotted straight lines, at high temperatures both
$1/L_{g}(T)$ and $R_{z}(T)$ exhibit an Arrhenius-like behavior, thereby
pointing to a thermally activated process involving single vortices. In
this regime $R_{z}(T)$ is almost frequency independent, whereas
$1/L_{g}(T)$, as shown in the insert of Fig. 1, is proportional to
$\omega^{2}$. These features are in excellent agreement with
what one expects from barrier limited diffusion of non interacting particles. 

The linearity of the Arrhenius plots in the thermally activated
regime implies that the activation energy $U(T,B)$ is either
constant or linearly dependent on temperature.
We expect the latter to be the case, since $U(T,B)$ should be proportional 
to the basic energy scale of vortex matter, 
$\epsilon(T)=\phi_{0}^{2}/4\pi\mu_{0}\lambda_{ab}^{2}(T)$($\phi_{0}$ is 
the superconducting flux quantum and $\mu_{0}$ the induction
constant), which varies as $(T_{c}-T)$ in the temperature range of interest
just below $T_{c}$. Then, 
noticing that the temperature dependence of $\lambda_{ab}^{-2}$ discussed above
leads to $U(T,B)\approx 2U(0,B)[1-(T/T_{c})]$ 
near $T_c$, 
our data allow to explore the dimensionality of the vortex
medium by studying the field dependence of the zero-temperature activation 
energy $U(0,B)$. 

\begin{figure}
\centerline{\psfig{file=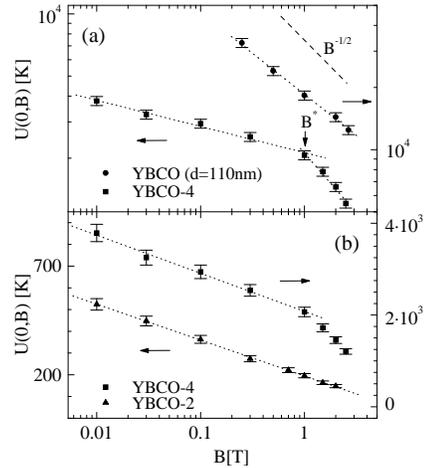,width=0.63\columnwidth,clip=}}
\vspace{1mm}
\caption{($a$) Log-log and ($b$) lin-log plots of the zero-temperature 
activation energy as a function of the magnetic field. The dashed lines are 
fits through the data. The dotted line in ($a$) is a $B^{-1/2}$ power 
law. $B^{*}$ indicates the 2D-3D crossover of the vortex medium in 
YBCO-4.}
\label{fig2}
\end{figure}

The results for the three YBCO samples studied in this work are shown
in Fig. 2. 
While the energy barrier
for the thickest (reference) film [Fig. 2(a)] obeys a power law
$U(0,B)\propto B^{-\alpha}$ with an exponent $\alpha\approx 0.40\pm 0.05$
fairly close to the prediction $\alpha=1/2$ for plastic vortex motion in
a 3D vortex liquid \cite{Rev}, for the thinnest YBCO-2 layer [Fig.
2(b)] $U(0,B)$ exhibits, over the entire field range covered by our experiments,
a logarithmic field dependence signaling 2D behavior. This interpretation
is corroborated by a quantitative comparison with the theoretical predictions
for flux flow in two dimensions controlled by surface barriers \cite{B}.
In this regime the prelogarithmic factor of $U(0,B)$ is of the form
$C\epsilon(0)d$, with $C=1/2$. Inspection of the slope of $U(0,B)$ for 
YBCO-2 in Fig. 2(b) gives $\partial U(0,B)/\partial\ln(1/B)\approx 71K$,
corresponding to $C\approx 0.45$, a value in excellent agreement with that 
calculated for the surface barrier mechanism. 
Note that the response of samples of macroscopic size as those studied 
in this work would be dominated by edge barriers only at extremely low 
frequencies, far below those accessible to our experiment. However, as 
evidenced by AFM images of the film microstructure, a consequence of the 
"unit cell by unit cell" growth of the YBCO layers is the formation of 
linear defects (steps), related to thickness variations in multiples of 
a unit cell, separating flat islands $\sim 0.2\mu m$ in size. 
It can be then expected that such steps play a role similar to that of 
sample edges, as they would also provide barriers against vortex motion 
exhibiting, in two dimensions, a logarithmic field dependence. 

Additional evidence for the 2D-3D dimensional crossover of
the vortex medium is provided by a study of the activation
energy for YBCO-4. As shown in Fig. 2(a), for this film $U(0,B)$
exhibits
algebraic behavior with $\alpha\approx 0.58\pm 0.06$ at high fields, but
crosses over, at $B^{*}\approx 1T$, to a low-field regime characterized
by a much smaller exponent ($\alpha\approx 0.1$) pointing to a logarithmic
field dependence, which is indeed demonstrated by the lin-log plot of Fig.
2(b). By comparing the activation energies for 3D \cite{Rev,B}
and 2D \cite{B} vortex liquids, it is possible to estimate
the crossover field as $B^{*}\approx k\phi_{0}(\gamma/d)^{2}$, where
$\gamma$ is the anisotropy ratio and $k$ a numerical constant of order unity. Using
$\gamma\approx 1/7$ for YBCO, one obtains $B^{*}\approx 1T$ for
YBCO-4 by choosing $k\approx 1/2$.

As shown by Fig. 1, with decreasing temperature both $1/L_{g}(T)$ and $R_{z}(T)$
cross over to a regime where their temperature dependence becomes much
weaker than in the activated regime. While the change in behavior
in $1/L_{g}(T)$ would be present even if $R_{z}(T)$ would continue to decrease
exponentially with $1/T$, the crossover in $R_{z}(T)$ points to the onset
of a different regime. In order to elucidate its physical
nature, it is convenient
to define a crossover temperature $T^{*}(\omega)$ and to compare it with
the melting temperature of the vortex medium. We \textit{ad hoc} identify
$T^{*}(\omega)$ as the temperature corresponding to the maximum curvature
of the $R_{z}(T)$-curves in Fig. 1, the particular choice of the criterion 
being irrelevant for our conclusions. At $B=0.01T$, 
$T^{*}(\omega)$ varies approximately from $67K$ at the upper limit to $64K$
at the lower end of the frequency spectrum explored in our measurements
(the maximum curvature in the $1/L_{g}(T)$-curves occurs at about the same
temperatures). Quite remarkably, these crossover temperatures are much
higher than the temperature,
$T_{M}=\phi_{0}^{2}d/32\pi^{2}\sqrt{3}k_{B}\mu_{0}\lambda_{ab}^{2}(T_{M})\approx15K$
($k_{B}$ is the Boltzmann constant), at which the 2D vortex crystal
in YBCO-4 would melt due to dislocation unbinding \cite{HD}
in the absence of disorder
(notice that the data of
Fig. 1 were taken in the 2D regime well below $B^{*}$). Thus, the picture
emerging from the impedance measurements is that of a vortex liquid which,
well below $T^{*}(\omega)$, appears to be frozen at the time scales,
$1/\omega$, of our measurements and whose response, as is shown below, can 
be described by ac extensions of the theory of collective pinning.

The low-frequency linear dynamic response of a collectively pinned elastic
vortex manifold has been investigated in Refs. \cite{KV,FFH} and
recently, in a   more systematic way, in Ref. \cite{K}. In this
approach large (in general anisotropic) portions of the vortex medium
(vortex bundles) are assumed to be fluctuating 
between pairs of metastable states in the uncorrelated random potential.
Treating these pairs of states as current-driven two-level systems with 
a size distribution $\nu(r)$,
the vortex contribution $l_{zv}(T,\omega)$ to the specific
inductance of the system can be shown \cite{KV,K} to be, quite
generally, of the form: 
\begin{equation} l_{zv}(T,\omega)\sim
B^{2}\int_{r_{c}}^{r_{\omega}}\,dr\nu(r)\frac{V^2u^2}{E}
\label{eq1}
\end{equation}
where $u$ is the average vortex displacement inside a bundle in the direction 
of the current-induced force, $r$ the size of a bundle in the same 
direction, $V$ its volume, $E$ the energy scale which can be 
associated with it,
$r_{c}$ the collective pinning length \cite{LO} and $r_{\omega}\sim 
r_{c}[(k_{B}T/U_0)\ln(1/\omega\tau)]^{1/\chi}$ a frequency-dependent length 
scale setting the maximum size of the vortex bundles which appreciably 
contribute to the response ($U_0$ and $\tau$ are, respectively, a 
characteristic energy and a relaxation time related to disorder, 
$\chi$ the energy barrier exponent).
It turns out to be convenient to estimate $E$ by
considering the compressive contribution $E_{c}$ to the energy.
In doing this, one has to take into account the dispersive
nature of the compression modulus $c_{11}(q)$, which in thin films
($d\ll\lambda_{ab}$) is always nonlocal.
Using $c_{11}(q)\approx(B^{2}d/\mu_{0}\lambda_{ab}^{2})q^{-2}$ in
the regime of interest, one obtains \cite{Rev}
$E_{c}\sim(B^{2}d/\mu_{0}\lambda_{ab}^{2})S^{2}(u/r)^{2}$, where 
$S=V/d\propto rr_{\perp}$.
Then, setting $\nu(r)\propto 1/r^3$
as imposed by the presence of a hierarchical distribution of 
quasi-isotropic ($r_{\perp}\propto r$) two-level
systems \cite{K}, 
from Eq. (1) we find: 
\begin{equation}
L_{z}(T,\omega)\approx
L_{0}(T)\{1+C^{'}\ln[(k_{B}T/U_0)\ln(1/\omega\tau)]\},
\label{eq2}
\end{equation}
where $C^{'}$ is a numerical constant. Since $L_{z}(\omega)$
depends only logarithmically on $\omega$, a simplified Kramers-Kronig 
relation \cite{PI} can be used which leads to:
\begin{equation}
R_{z}(T,\omega)\approx C^{'}(\pi/2)\omega L_{0}(T)/\ln(1/\omega\tau).
\label{eq3}
\end{equation}
Notice that in a film with a high density of linear defects (steps) acting 
as a network of strong pinning lines, the regime of collective pinning 
can be realized by vortices moving along these linear defects. 

The main features of the response of our very thin YBCO
layers in the frozen vortex-liquid regime below $T^{*}(\omega)$
are well described by these expressions (note that, below $T^{*}(\omega)$, 
$R_{z}\ll\omega L_{z}$ and, therefore, $L_{g}\approx L_{z}$). Focusing 
first on the
temperature dependence, one sees that it should be dominated by
$L_{0}(T)$, the logarithmic corrections entering Eqs. (2) and (3)
varying too slowly to be of any relevance in the limited
temperature interval of our experiments. The dashed curve in Fig.
1(a), which is simply $1/L_{0}(T)$ as inferred from zero-field 
measurements, fits nicely the low-temperature (almost
frequency-independent) $1/L_{g}(T)$-curves. On the other hand, the 
temperature dependence and the order of magnitude of $R_{z}$ below 
$T^{*}(\omega)$ are compatible with Eq. (3) for any reasonable 
estimate of $\tau$ ($10^{-12}-10^{-7} s$). 
We have also found that the response in the 
frozen vortex-liquid regime is only weakly dependent on $B$ in the field range
covered by our experiments (up to $3T$), in agreement with the 
theoretical prediction (derived for $B\ll B_{c2}$).

\begin{figure}
\centerline{\psfig{file=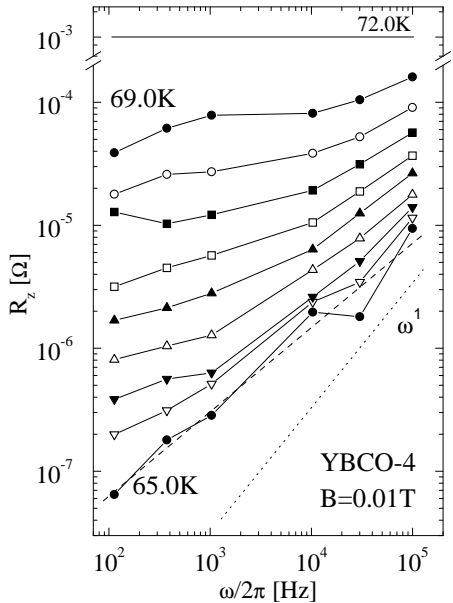,width=0.68\columnwidth,clip=}}
\vspace{1mm}
\caption{Sheet resistance vs frequency isotherms taken at $0.5K$-intervals 
between $65K$ and $69K$ for YBCO-4 at $B=0.01T$ on a log-log plot. The 
dashed line is a power-law fit of the $65K$-data with an exponent 
$\approx 0.7$. The dotted line shows the (almost) linear frequency dependence 
predicted by Eq. (3).}
\label{fig3}
\end{figure}

Further evidence for the interpretation of the response in terms of a
collectively pinned vortex manifold emerges from the analysis of the frequency
dependence of our data. As shown in Fig. 1(a), at low temperatures, well
below $T^{*}(\omega)$, $1/L_{g}$ is almost frequency independent, a behavior
consistent with Eq. (2), where the extremely slow-varying (double-logarithmic)
vortex contribution can be hardly expected to be noticeable against the 
superfluidbackground $1/L_{0}$. To discuss the dissipative component, in 
Fig. 3 we show a family of $R_{z}$ vs $\omega$ isotherms in a log-log plot. With
decreasing temperature the isotherms progressively evolve from the
frequency-independent regime characteristic of the vortex liquid at high
temperatures to an almost algebraic behavior with an exponent $\sim
0.7$ at the lowest temperature ($65K$) at which dissipation could be studied
with sufficient accuracy. Considering the fact that this isotherm reflects,
at a time scale $\sim 1/\omega$, the response of a 2D vortex medium "on
the verge of freezing" rather than that of a "deep-frozen" liquid, we interpret
the general behavior emerging from Fig. 3 and, in particular, the value
of the exponent extracted from the "coldest" isotherm, as an indication
that $R_{z}(\omega)$ will likely tend to the almost linear frequency dependence
predicted by Eq. (3) well below $T^{*}(\omega)$. This should be compared
with the response of a non-frozen pinned vortex liquid for which one would 
expect a crossover to an $\omega^2$-dependence at sufficiently low temperatures. 


We would like to thank L. Baselgia Stahel, S. Blaser and B.
Schmied for their contribution during the early stages of this
work, A. Daridon for the AFM images and A. R\"{u}fenacht for assistance 
in the analyis of the data. We are also grateful to G. Blatter and V.B. 
Geshkenbein for interesting discussions and useful comments. This work was
supported by the Swiss National Science Foundation.


%
 

\end{document}